\input psfig
\documentstyle[twoside,fleqn,espcrc2]{article}

\newcommand{\AmS}{{\protect\the\textfont2
  A\kern-.1667em\lower.5ex\hbox{M}\kern-.125emS}}

\title{Hadronic Correlators from All-point Quark Propagators}
\author{ A.~Duncan\thanks{Talk presented by A.Duncan}\address{Dept. of Physics and Astronomy,University of Pittsburgh,
 Pittsburgh, PA 15260},%
  E.~Eichten\address{ Theory Group, Fermilab, PO Box 500, Batavia, IL60510},%
  and 
  J.~Yoo $^a$}
\begin{document}
\begin{abstract}
A method for computing all-point quark propagators is applied to a variety of
processes of physical interest in lattice QCD. The method allows, for example,
efficient calculation of disconnected parts and full momentum-space 2 and
3 point functions. Examples discussed include: extraction of chiral Lagrangian
parameters from current correlators, the pion form factor, and the unquenched
eta-prime.
\end{abstract}

\maketitle

\section{Methodology}

 The very high cost of generating decorrelated dynamical gauge configurations
makes it increasingly important to extract the maximum physical information content
of each available configuration. In many cases, this requires calculation of
hadronic observables involving quark propagators from any point on the lattice to any
other. Here we describe an approach to obtaining such propagators by simulating
bosonic pseudofermion fields. Introduce bosonic pseudofermion field $\phi_{ma}$ with action ($m$ a lattice site, $a$ the spin-color index, $Q$ the Wilson or clover operator):
\begin{eqnarray*}
   S(\phi)&=& \phi^{\dagger}Q^{\dagger}Q\phi \\
          &=& \phi^{\dagger}H^2 \phi,\;\;\;H \equiv \gamma_{5}Q = H^{\dagger}
\end{eqnarray*}
For fixed background gauge field $A$, simulating the pseudofermion field with the
preceding action produces the following correlator ($<<O>>$ means the average of
 $O$ relative to the measure $e^{-S}$)  :
\begin{eqnarray*}
   <<\phi_{ma}\phi^{*}_{nb}>>_{S(\phi)}&=& (H^{-2})_{ma,nb} \\
   <<\phi_{ma}(\phi^{\dagger}H)_{nb}>>_{S(\phi)} &=& (H^{-1})_{ma,nb}\\           
   &=&(Q^{-1}\gamma_{5})_{ma,nb}
\end{eqnarray*}
These simulations are practical for two reasons:\\
(i) The pseudofermion average $<<....>>$  is efficiently implemented by heat-bath update of pseudofermion fields.\\
(ii) For a fixed gauge field, most quantities decorrelate after a few pseudofermion sweeps.\\
 The computation of multipoint hadronic correlators involving $n$ quark propagators
 can be reduced to convolutions of $n$ pseudofermion fields, rapidly computed by 
 fast Fourier transform (FFT). For example, the full 4-momentum transform 
 $\Delta(q)\equiv\sum_{x,y}e^{iq\cdot(x-y)}\Delta(x,y)$ of  the 2-point
pseudoscalar correlator 
\begin{eqnarray*}
 \Delta(x,y)&=& <0|T\{\bar{\Psi}(x)\gamma_{5}\Psi(x)\;\bar{\Psi}(y)\gamma_{5}\Psi(y)\}|0> \\
            &=& -<\rm{tr}((Q^{-1}\gamma_{5})_{xy}(Q^{-1}\gamma_{5})_{yx})> \\
            &=& -<<\sum_{ab}\phi_{xa}(\phi^{\dagger}H)_{yb}\chi_{yb}(\chi^{\dagger}H)_{xa}>> \\
            &=& -<<(\phi^{\dagger}H\chi)_{yy}(\chi^{\dagger}H\phi)_{xx}>>
\end{eqnarray*}
becomes
\begin{eqnarray*}
\Delta(q)   = -<< \rm{FFT}(\chi^{\dagger}H\phi)(q)\rm{FFT}(\phi^{\dagger}H\chi)(-q)>>
\end{eqnarray*}

\section{Applications}
 We have studied the feasibility of a pseudofermion approach to all-point propagators
in a number of examples of physical interest. A brief report on the progress to date
follows.
\subsection{Chiral Lagrangian Parameters from Current Correlators}
 The chiral Lagrangian predicts the  low-momentum structure of 2-point functions
 such as $\Delta(q)$ in QCD \cite{Gass}:
\begin{eqnarray*}
   \Delta(q) &\simeq& \frac{G_{\pi}^2}{q^2+M_{\pi}^{2}}+\frac{B^2}{2\pi^2}(h_1-l_4)+Cq^2+O(q^4) \\
   G_{\pi} &=& \frac{F_{\pi}M_{\pi}^2}{m_q}  \\
   BF_{\pi}^2 &=& <\bar{\Psi}\Psi> 
\end{eqnarray*}
  We have studied various correlators of this type using  800 dynamical configurations 
generated using the truncated determinant (TDA) algorithm \cite{TDA} on large coarse (6$^4$)
lattices. 
The presence of low eigenmodes of $H$ leads to longer autocorrelation times for the zero
momentum component (see Fig.1), on the order of 60 pseudofermion sweeps, with much
more rapid decorrelation for nonzero momentum values of $\Delta(q^2)$. 
 (Low momentum autocorrelations can be substantially reduced
by projecting out the lowest few eigenmodes of $H$- the required code has been written
and is presently being tested \cite{DuncEich}). However, even at zero momentum, the intrinsic gauge
 fluctuations exceed the statistical errors from the pseudofermion evaluation (which become
 very small at higher momenta where decorrelation is rapid). This can be seen in Fig.2,
 where we display the measured $\Delta(q^2)$ for two separate configurations, as well
 as the average over  800 configurations and a fit to the chiral formula. In fact a good fit to the predicted chiral form can be obtained
even without the zero-momentum point. 
\begin{figure}
\psfig{figure=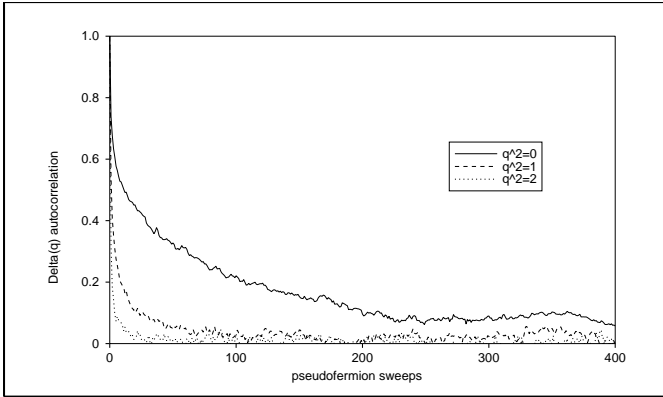,height=0.7\hsize}
\vspace*{-0.45in}
\caption{Decorrelation of pseudofermion evaluations of $\Delta(q)$}
\end{figure}
The parameters $h_1,l_4$ are couplings in the next-to-leading  order chiral Lagrangian \cite{Gass}.
 Fitting the measured $\Delta(q)$ (800 6$^4$ unquenched TDA lattices  \cite{eichten} with O($a^2$) improved
 gauge action, $a$=0.4 F):
\begin{eqnarray*}
  \Delta(q) = \frac{A_1}{q^2+A_2}+A_3+A_4 q^{2}+A_5 (q^{2})^{2}
\end{eqnarray*}
A standard cosh fit of smeared-local correlators gives $M_{\pi}=0.396 \pm 0.007$. 
Allowing the pion mass to vary in the chiral formula, the best fit  (see Fig.2) is obtained in the range 0.25$<q^2<$2.5 GeV$^2$, and gives a 2\% evaluation  (jackknife errors)
of the one-loop chiral parameter
$A_3$:
\begin{eqnarray*}
    A_{1}&=& 23.1 \pm 0.4 \\
    A_{2}&=& 0.178 \pm 0.017 \Rightarrow M_{\pi}=0.422\pm0.020\\
    A_{3}&=& 9.48 \pm 0.16 \\
    A_{4}&=& -0.70 \pm 0.02,\;\; A_{5}= 0.025\pm 0.001 \\
\end{eqnarray*}
\begin{figure}
\psfig{figure=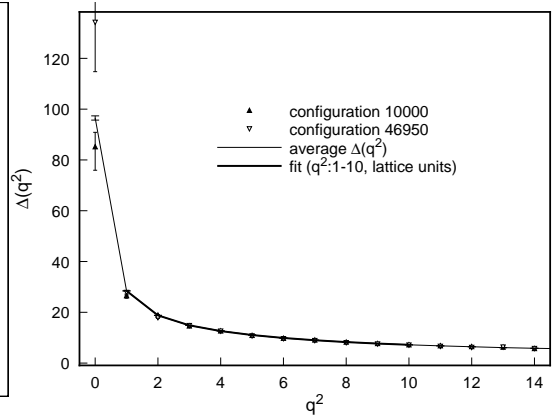,height=0.75\hsize}
\vspace*{-0.45in}
\caption{4 parameter fit of measured $\Delta(q^2)$}
\end{figure}
\vspace*{-0.45in}
\subsection{Three-point Functions: the Pion Form Factor}
To extract the pion formfactor, we need the following 3-point function (see Fig.3):
\begin{eqnarray*}
J_{t_{0}t_{1}t_{2}}(\vec{q}^2) &=& \sum_{\vec{w}\vec{x}\vec{y}\vec{z}}e^{i\vec{q}\cdot(\vec{x}-\vec{y})}f^{\rm sm}(\vec{z})f^{\rm sm}(\vec{w})\\
&&\hspace{-0.4in}<\bar{\Psi}(\vec{z}+\vec{x},t_2)\gamma_5\Psi(\vec{x},t_2)\bar{\Psi}(\vec{y},t_1)\gamma_{0}\Psi(\vec{y},t_1)\\
&&\hspace{-0.4in}\bar{\Psi}(\vec{w},t_0)\gamma_5\Psi(0,t_0)>
\end{eqnarray*}
\begin{figure}
\psfig{figure=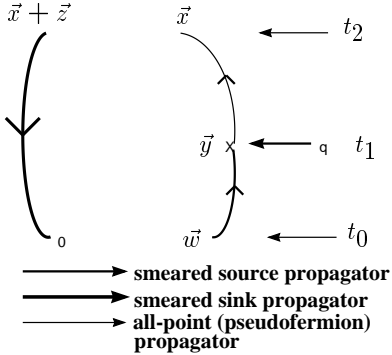,height=0.65\hsize}
\vspace*{-0.41in}
\caption{3-point function  for pion form factor}
\end{figure}
 To evaluate $J(q^2)$ simultaneously for all spacelike $q$ injected at time $t_1$ by the
 electromagnetic current, it suffices to use a conventional smeared source propagator
 for the quark propagation from $\vec{w},t_0$ to $\vec{y},t_1$ and a smeared sink
propagator for the propagation from $\vec{x}+\vec{z},t_2$ to $\vec{0},t_0$. The
 remaining propagator from $\vec{y},t_1$ to $\vec{x},t_2$ is then needed for all
 source and sink points (to obtain the full momentum space Fourier transform) and
 is evaluated by the pseudofermion technique. At this stage, both the connected and
 disconnected contributions (where the quark propagates from  $\vec{y}$ back to $\vec{y}$)
 to $J(q^2)$ are readily available, as the all-point 
propagator also gives us the amplitude for all coincident source and sink points.
 An approximate pion form factor is given by $J(q^2)e^{(t_2-t_1)E(q)}/E(q))$ ($t_0,t_1,t_2$=0,3,6, with $E(q)$ the energy for
 a lattice pion of  momentum $q$): results 
 obtained with 60 quenched 12$^3$x24 lattices at $\beta=$5.9 are  shown in Fig. 4.
Analysis of a single gauge configuration takes about 12 hours on a Pentium 4 processor.
 Reliable calculation of the pion form factor requires use of optimized  smearing wavefunctions $f^{\rm sm}$ to project 
 out ground-state pions as the signal dies quickly at larger Euclidean times $t_1-t_0,t_2-t_1$,
 especially at larger momentum, and to check that a plateau is reached at large
Euclidean time.  Of  course, more accurate results require 
  higher statistics than used in this  feasibility study.
\vspace{-0.05in}
\begin{figure}
\psfig{figure=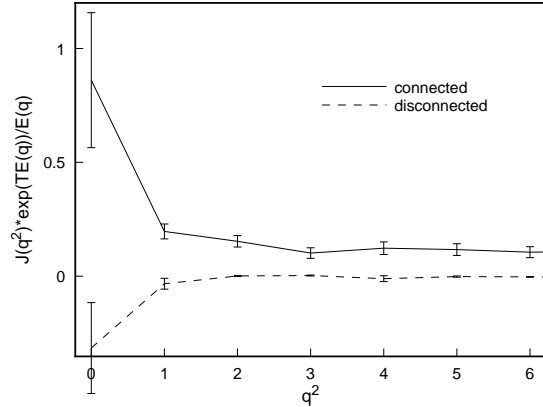,height=0.75\hsize}
\vspace*{-0.5in}
\caption{Connected and Disconnected contributions to Pion Form Factor}
\end{figure}
\subsection{The unquenched eta-prime}
 We may extract the  etaprime propagator in the isoscalar  channel using two pseudofermion
fields to generate propagators for both quark lines:
\begin{eqnarray*}
\Delta_{\eta}(p)&= &\sum_{xy\vec{z}\vec{w}}e^{ip\cdot(x-y)}f^{\rm sm}(\vec{z})f^{\rm sm}(\vec{w})\times\\
 &&\hspace{-0.3in}<\bar{\Psi}(\vec{z}+\vec{x},x_{4})\gamma_{5}\Psi(x)\bar{\Psi}(\vec{w}+\vec{y},y_4)\gamma_{5}\Psi(y)>  
\end{eqnarray*}
 In terms of pseudofermion averages, this gives a connected
\begin{eqnarray*}
&&\hspace{-0.3in} \sum_{xy\vec{z}\vec{w}}e^{ip\cdot(x-y)}f^{\rm sm}(\vec{z})f^{\rm sm}(\vec{w})\\
 &&<<\phi(x)(\phi^{\dagger}H)(y+\vec{w})\chi(y)(\chi^{\dagger}H)(x+\vec{z})>>
\end{eqnarray*}
as well as disconnected (``hairpin") contribution (for two flavors of sea quarks):
\begin{eqnarray*}
&&\hspace{-0.3in} -2\sum_{xy\vec{z}\vec{w}}e^{ip\cdot(x-y)}f^{\rm sm}(\vec{z})f^{\rm sm}(\vec{w})\\
 &&<<\phi(x)(\phi^{\dagger}H)(x+\vec{z})\chi(y)(\chi^{\dagger}H)(y+\vec{w})>> 
\end{eqnarray*}
Computations of the unquenched eta-prime propagator using this method are in
progress, using 10$^3$x20 lattices generated with the TDA algorithm.

\vspace{0.2in}



\end{document}